\magnification=\magstep1
\hoffset=0.1truecm
\voffset=0.1truecm
\vsize=23.0truecm
\hsize=16.25truecm
\parskip=0.2truecm
\def\pp{\parshape 2 0.0truecm 16.25truecm 2truecm 14.25truecm}
\def\sigbar{ {\langle \sigma \rangle}}
\def\alphabar{ { {\bar \alpha} }}
\def\gnu{\Gamma_\nu}

\def\con{{\Lambda}} 
\def\fraclm{{ {\cal F}_1 }}
\def\frachm{{ {\cal F}_2 }}
\def\newpage{\vfill\eject}
%
%
%
\centerline{\bf IMPLICATIONS OF WHITE DWARF GALACTIC HALOS} 
\bigskip
\centerline{\bf Fred C. Adams and Gregory Laughlin}
\bigskip
\centerline{\it Physics Department, University of Michigan}
\centerline{\it Ann Arbor, MI 48109, USA}
\medskip 
\centerline{fca@umich.edu \quad and \quad gpl@umich.edu} 
\vskip 0.15truein
\centerline{\it submitted to The Astrophysical Journal} 
\medskip 
\centerline{1 February 1996} 
\vskip 0.4truein 

\bigskip 
\centerline{\bf Abstract} 
\medskip 

Motivated by recent measurements which suggest that roughly half the
mass of the galactic halo may be in the form of white dwarfs, we study
the implications of such a halo.  We first use current limits on the
infrared background light and the galactic metallicity to constrain
the allowed initial mass function (IMF) of the stellar population that
produced the white dwarfs.  The IMF must be sharply peaked about a
characteristic mass scale $M_C \approx 2.3 M_\odot$.  Since only a
fraction of the initial mass of a star is incorporated into the
remnant white dwarf, we argue that the mass fraction of white dwarfs
in the halo is likely to be 25\% or less, and that 50\% is an extreme
upper limit.  We use the IMF results to place corresponding
constraints on the primordial initial conditions for star formation.
The initial conditions must be much more homogeneous and skewed toward
higher temperatures ($T_{\rm gas} \sim$ 200 K) than the conditions
which lead to the present day IMF.  Next we determine the luminosity
function of white dwarfs.  By comparing this result with the observed
luminosity function, we find that the age of the halo population must
be greater than $\sim 16$ Gyr.  Finally, we calculate the radiative
signature of a white dwarf halo.  This infrared background is very
faint, but is potentially detectable with future observations.

\vskip 0.4truein 

\noindent
{\it Subject Headings:} dark matter -- galaxies: structure -- 
stars: white dwarfs -- stars: evolution -- stars: formation 

\newpage 
\centerline{\bf 1. INTRODUCTION} 
\medskip 

The nature of the dark matter that makes up galactic halos is an
important unresolved astrophysical issue.  Microlensing experiments
(Alcock et al. 1993; Aubourg et al.  1993) have indicated the presence
of some type of low mass stellar objects in our galactic halo.  Recent
measurements (MACHO collaboration 1996; see also Bennet et al. 1996) 
suggest that a substantial fraction (roughly half) of the halo mass is
composed of white dwarfs, the remnants of an early generation of
stars.  In this paper, we examine the implications of a galactic halo
filled with white dwarfs.  We first determine necessary constraints on
the distribution of masses for the stellar generation that produced
these white dwarfs. We show that the resulting initial mass function
(IMF) is very different from the present day IMF.  In addition, its
highly peaked form provides remarkable constraints on the initial
conditions for star formation.  We then show how a population of halo
white dwarfs affects the observed luminosity function of white dwarfs.
We find that in order to be consistent with the observed luminosity
function of white dwarfs, the halo population is likely to have an age
$\sim 16$ Gyr.  We then use our synthesized luminosity functions to
determine the radiative signature of the halo.

The idea that white dwarfs and other stellar remnants 
(e.g., neutron stars) are present in large quantities 
in galactic halos has been considered by several previous authors
(e.g., Hegyi \& Olive 1983, 1986, 1989; Ryu, Olive, \& Silk 1990). 
Neutron stars are essentially ruled out because their progenitors 
are massive stars which leave behind too much mass in the 
form of heavy elements when they explode in supernovae. 
White dwarfs can be a viable candidate for the halo dark matter 
provided that the initial mass function (IMF) of the progenitor 
stars is confined to a narrow mass range, roughly $1 < m < 8$.
Throughout this paper, we write stellar masses in solar units, 
i.e., we define $m \equiv M_\ast/(1 M_\odot)$.  Population 
synthesis models show that the bright early phases of these 
putative white dwarf halos can be detectable in deep galaxy 
counts and hence are further constrained (Charlot \& Silk 1995). 

The logic and organization of this paper can be summarized as
follows. Using the idea that white dwarfs comprise a substantial
fraction of the present day galactic halo, we find constraints on the
initial mass function at the epoch of star formation in the halo 
(\S 2).  This result is then used in conjunction with current theories
of the IMF to constrain the physical conditions that led to star
formation (\S 3).  Next, we calculate the luminosity function of 
this white dwarf population (\S 4); we show that if sufficiently 
sensitive measurements of this luminosity function can be made, 
then the age of the halo can be cleanly determined. 
We determine the expected background radiation field from this
galactic halo (\S 5). Finally, we conclude (\S 6) with a summary 
and discussion of our results. 

\newpage 
\centerline{\bf 2. THE IMPLIED IMF OF THE FIRST STELLAR GENERATION} 
\medskip 

In this section, we constrain the IMF of the stellar generation that
produced the lensing white dwarfs observed in the galactic halo.  For
simplicity, we assume that most of the stars in the halo were produced
in a single burst of star formation that occurred some time $\tau_H$
in the past.  We show that the allowed IMF for this stellar
distribution is highly constrained.  Since current observations have
placed tight limits on the mass fraction of small stars (red dwarfs)
in the galactic halo, the IMF is constrained on the low mass end 
($m < 1$).  As we show below, the mass fraction of high mass stars 
($m < 8$) is also highly constrained because these stars end their lives
in supernova explosions and thereby contaminate the interstellar
medium with heavy elements.  The net result is that if an initial
stellar population produced white dwarfs which are currently a major
constituent of the galactic halo, then the IMF must be rather
tightly confined to the mass range $1 < m < 8$.  The remainder of
this section is devoted to quantifying this assertion.

In order to proceed quantitatively, we require a description of the
IMF.  For the sake of definiteness, we consider the distribution
$f=dN/d\ln m$ of stellar masses to be of the general log-normal form 
$$\ln f (\ln m) = A - {1 \over 2 \sigbar^2}
\Bigl\{ \ln \bigl[ m / m_C \bigr] \Bigr\}^2 \, , \eqno(2.1)$$
where $A$, $m_C$, and $\sigbar$ are constants. This general form 
for the IMF is motivated by the current theory of star formation 
and by general statistical considerations 
(Adams \& Fatuzzo 1996; see also \S 3; Zinnecker 1984, 1985; Larson 1973; 
Elmegreen \& Mathieu 1983). This form for the IMF also has sufficient 
flexibility to assume a wide variety of behavior. The parameter $A$ 
determines the overall normalization of the distribution; the parameter 
$m_C$ is the mass scale (given here in solar units) which sets the 
center of the distribution; the parameter $\sigbar$ is the dimensionless 
width of the distribution.  Notice that the shape of the
distribution is completely determined by the mass scale $m_C$ 
and the total width $\sigbar$. As a reference point, we note that 
if the present day IMF is fit with a log-normal form, then the 
shape parameters have the values $\sigbar \approx$ 1.57 and 
$m_C = 0.1 - 0.2$ (see Miller \& Scalo 1979; Scalo 1986; 
Adams \& Fatuzzo 1996).  As we show below, these shape parameters 
are highly constrained for the stellar population that filled the 
galactic halo with white dwarfs. 

Since the age of the galaxy is $\sim$10 -- 20 Gyr, only those stars in 
the halo with $m > 1$ have had time to evolve into white dwarfs.  Stars
with smaller masses ($m < 1$) are still burning hydrogen and
contributing to the infrared background light.  Here we define the
mass fraction $\fraclm$ of the original stellar population in low mass
stars to be $\fraclm \equiv M_{RD} / M_{TOT}$, where $M_{RD}$ is the mass
incorporated into stars in the range $m < 1$ and $M_{TOT}$ is the total 
mass of the initial stellar population. Using the form (2.1) for the mass 
distribution, we can write this mass fraction in the form 
$$\fraclm = {M_{RD} \over M_{TOT} } = {1 \over 2} \Bigl\{ 
1 - {\rm Erf} (\xi_1) \Bigr\} \, , \eqno(2.2)$$
where ${\rm Erf}(\xi)$ is the error function (see Abramowitz \& 
Stegun 1970) and where the value $\xi_1$ is related to the parameters 
in the IMF, 
$$\xi_1 \equiv { \sqrt{2} \over 2} \Bigl\{ {\ln m_c \over \sigbar} 
+ \sigbar \Bigr\} \, . \eqno(2.3)$$
For a given mass fraction $\fraclm$, we obtain a constraint on the 
parameters in the IMF:  
$$\ln m_c + \sigbar^2 = \sigbar \sqrt{2} \, 
{\rm Erf}^{-1} [1 - 2 \fraclm] \, . \eqno(2.4)$$ 

Similarly, the mass fraction of high mass stars is limited by 
metallicity considerations.  We define this mass fraction to be 
$\frachm \equiv M_{HM} / M_{TOT}$, where $M_{HM}$ is the total mass 
of the initial stellar population in high mass stars with $m > 8$. 
The mass fraction in high mass stars can be written 
$$\frachm = {M_{HM} \over M_{TOT} } = {1 \over 2} \Bigl\{ 
1 - {\rm Erf} (\xi_2) \Bigr\} \, , \eqno(2.5)$$
where $\xi_2$ is also related to the IMF parameters 
and is given by 
$$\xi_2 \equiv { \sqrt{2} \over 2} \Bigl\{ 
{\ln (8/m_c) \over \sigbar} - \sigbar \Bigr\} \, . \eqno(2.6)$$
Thus, for a given mass fraction $\frachm$ in high mass stars, 
we obtain a second constraint on the parameters in the IMF: 
$$\ln (8/m_c) - \sigbar^2 = \sigbar \sqrt{2} \,  
{\rm Erf}^{-1} [1 - 2 \frachm] \, . \eqno(2.7)$$ 

For given values of the mass fractions $\fraclm$ and $\frachm$, 
equations (2.4) and (2.7) specify the two unknown quantities 
in the initial IMF.  We must thus estimate the mass fractions 
in both low mass and high mass stars. 
Recent work has shown that faint red stars do not contribute
significantly to the mass budget of the galactic halo (Bahcall 
et al. 1994; Graff \& Freese 1996).  This work implies that 
we can take $\fraclm \approx 0.01$ as a fairly conservative estimate. 
We can obtain a rough estimate of the high mass fraction $\frachm$ 
from a simple metallicity argument. The amount of metals (heavy 
elements) produced by the high mass stars in the halo is given by 
$\Omega_{\rm Metal} = \Omega_{\rm Halo} \frachm f_Z$, where $f_Z$ is the 
fraction of a high mass star that is ejected into the interstellar medium 
in the form of metals, $\Omega_{\rm Halo}$ is the total mass density 
in the halo relative to the critical density of the universe, and 
$\Omega_{\rm Metal}$ is the relative fraction of metals. 
The metallicity $Z$ of the galactic disk is thus given by 
$$Z = {\Omega_{\rm Metal} \over \Omega_{\rm Disk} } = 
{\Omega_{\rm Halo} \over \Omega_{\rm Disk} } \frachm f_Z \, . \eqno(2.8)$$ 
If we take the fraction of ejected metals to be $f_Z = 0.1$ 
and $\Omega_{\rm Halo}/\Omega_{\rm Disk}$ = 10, then $\frachm \approx Z$. 
As a conservative limit, we can thus take $\frachm < 0.01$ 
(this rough argument is in good agreement with the previous 
results of Ryu et al. 1990; see also Hegyi \& Olive 1986). 
Using the representative values $\fraclm = 0.01 = \frachm$, we can 
evaluate the constraints (2.4) and (2.7) to obtain estimates 
for the shape parameters in the IMF, 
$$\sigbar = 0.44 \qquad {\rm and} \qquad  m_C = 2.3 \, . \eqno(2.9)$$ 
As expected, these values imply an IMF which is centered at a much
higher mass scale than the present day IMF (the mass scale $m_C$ is
larger by a factor of $\sim10$) and is much narrower (the width 
$\sigbar$ is smaller by a factor of $\sim3.5$).  The resulting IMF is
shown in Figure 1 (solid curve); a fit to the present day IMF
(consistent with the results of Miller \& Scalo 1979) is also shown
for comparison (dashed curve).  Notice that the IMF at the epoch of
star formation in the halo must be much more sharply peaked than that
of the present day.

The derived IMF shown in Figure 1 is the mass distribution that
saturates the constraints implied by equations (2.4) and (2.7).  We
can view this result another way by using the same constraints (this
time as inequalities rather than equalities) to define an allowed
region in the plane of parameters (i.e., the $m_C$-$\sigbar$ plane).
The result is shown in Figure 2. In the upper left corner of the
plane, we also show the point that corresponds to the parameters of
the present day IMF. Figures 1 and 2 underscore the fact that the IMF
of the halo population must be dramatically different from the present
day IMF.

In order to determine the distribution of masses for the white dwarf
population, we must specify the transformation between progenitor mass 
and white dwarf mass; unfortunately this relationship is somewhat uncertain 
(e.g., see Wood 1992 for further discussion 
of this issue).  For this paper, we use the following 
transformation between progenitor mass and white dwarf mass, 
$$m_{WD} = A_X \exp [ B_X m ] \, , \eqno(2.10)$$ 
with $A_X$ = 0.49 and $B_X$ = 0.095 (this formula is taken from
the standard model of Wood 1992).  For reference, we note that a
progenitor star with mass $m = m_C$ = 2.3 produces a white dwarf with
a mass $m_{WD} \approx 0.62$.  Using the IMF derived above, we can
calculate the mass function of white dwarfs in the halo.  The
resulting mass distribution of white dwarfs in the galactic halo is
shown in Figure 3. Notice that the distribution is sharply peaked
about $m_{WD} \approx 0.6$.  For comparison, the mass distribution of
white dwarfs resulting from the present day IMF is also shown.

Since only a fraction of the progenitor mass remains in the 
resulting white dwarf, there is an efficiency problem associated
with a galactic halo composed of white dwarfs.  For the sake of
definiteness, suppose that {\it all} of the mass in the galactic halo
is efficiently processed into a stellar population with an IMF $f =
dN/d\ln m$. We define the white dwarf efficiency ${\cal E}_{WD}$ to
be the mass fraction present in the white dwarfs resulting from the 
death of this initial stellar population. The value of ${\cal E}_{WD}$ 
depends on the IMF and the relationship between progenitor mass and 
white dwarf mass.  We can
write this efficiency in the form
$${\cal E}_{WD} = m_C^{-1} \, {\rm e}^{-\sigbar^2/2} \, 
{1 \over \sqrt{2 \pi} } \, \int_{z_1}^{z_8} \, dz \, 
{\rm e}^{-z^2/2} \, m_{WD}(z) \, , \eqno(2.11)$$
where the variable $z \equiv \ln(m/m_C)/\sigbar$ and 
where the limits of integration are given by the mass range 
which leads to white dwarf production:
$z_1 = z(m=1)$ and $z_8 = z(m=8)$. 
Using our derived IMF parameters (equation [2.9]) and the 
conversion formula (equation [2.10]), we 
obtain a white dwarf efficiency ${\cal E}_{WD}$ = 0.24. 
Thus, even if all of the halo was incorporated into an 
initial stellar population, only about 1/4 of the halo mass 
would remain in the form of white dwarfs (for our derived IMF). 

To obtain a higher white dwarf efficiency ${\cal E}_{WD}$, the IMF
must be tilted toward lower masses (since the function $m_{WD}/m$ is a
monotonically decreasing function of mass).  However, as the mass
scale $m_C$ in the IMF becomes smaller, the width $\sigbar$ must
also become smaller to avoid the overproduction of red dwarfs (see 
Figure 2). The limiting case is thus a delta function IMF at the mass
scale $m_C =1$.  In this limit, the white dwarf efficiency
(eq. [2.11]) reduces to ${\cal E}_{WD}$ = $m_{WD}/m$ = 0.54.  This
value represents the maximum allowed mass fraction of white dwarfs in
the halo.  Notice that this efficiency constraint can be avoided if
many stellar generations contribute to the halo population of white
dwarfs.  However, a solution involving multiple stellar generations is 
highly unlikely because it requires the same (very peculiar) IMF 
under rather different physical conditions. 

Another problem associated with this low white dwarf efficiency 
is that a large amount of gas is left over from the process. For 
example, if only 1/4 of the halo mass actually becomes incorporated 
into white dwarfs, then the remaining 3/4 of the halo mass must 
reside in some other type of baryonic dark matter.  Notice that only 
about 10\% of this material can be used to make up the current disk 
of the galaxy. Given the difficulties associated with baryonic 
dark matter (Hegyi \& Olive 1983, 1986), this problem is rather 
severe and makes the detection of large numbers of halo dwarfs 
(MACHO collaboration 1996) all the more startling.  

\bigskip 
\centerline{\bf 3. IMPLICATIONS FOR PRIMORDIAL INITIAL CONDITIONS} 
\medskip 

In this section, we wish to examine how the current theory of star
formation constrains the initial conditions during the epoch of star
formation in the galactic halo.  We thus require a relationship
between the IMF derived in the previous section and the initial
conditions. We use a recently developed theory of the IMF 
(Adams \& Fatuzzo 1996) to obtain this relation. 

Within the current paradigm of star formation (Shu, Adams, \& Lizano
1987), the process which determines the IMF can be divided into two
subprocesses: {\bf[1]} The spectrum of initial conditions produced by
the star forming environment.  {\bf[2]} The transformation between a
particular set of initial conditions and the properties of the final
(formed) star.  

In the current theory, the transformation {\bf [2]} is accomplished
through the action of stellar winds and outflows.  Stars are formed
through the collapse of centrally concentrated regions in molecular
clouds (cloud cores). The collapse produces a central star/disk system
at the center of the flow, with material falling onto the central
system at a well defined mass infall rate $\dot M$ $\sim$ $a^3/G$,
where $a$ is the effective sound speed (Shu 1977).  As a nascent star
gains mass, it becomes more luminous, and can produce an increasingly
more powerful stellar outflow.  When the strength of this outflow 
becomes larger than the ram pressure of the infalling material, the
star separates itself from the surrounding molecular environment and
thereby determines its final mass.

The transformation between the initial conditions and the final
stellar properties can be written as a ``semi-empirical mass formula''
(SEMF).  Using the idea that the stellar mass is determined when the
outflow strength exceeds the infall strength, we can write the SEMF in
the form  
$$L_\ast M_\ast^{2} = 8 m_0 \gamma^3 \delta
{\beta \over \alpha \epsilon}  {a^{11} \over G^3\Omega^2}
= \con \, {a^{11} \over G^3\Omega^2}  \, . \eqno(3.1)$$
This formula provides us with a transformation between
initial conditions (the sound speed $a$ and the rotation rate
$\Omega$) and the final properties of the star (the luminosity
$L_\ast$ and the mass $M_\ast$).  Furthermore, the protostellar 
luminosity as a function of mass is known so that equation (3.1) 
specifies the final stellar mass in terms of the initial conditions. 
In addition, the parameters $\alpha$, $\beta$, $\gamma$, $\delta$, and
$\epsilon$ are efficiency factors (see Shu, Lizano, \& Adams 1987; 
Adams \& Fatuzzo 1996).  In general, all of the quantities on the
right hand side of equation (3.1) will have a distribution of values.
These individual distributions ultimately determine the composite
distribution of stellar masses $M_\ast$.  However, as we argue below,
to leading order the mass distribution approaches a log-normal form.

In order to evaluate the semi-empirical mass formula (3.1), we must
specify the luminosity as a function of mass for young stellar
objects. In general, this luminosity has many contributions (Stahler,
Shu, \& Taam 1980; Adams \& Shu 1986; Adams 1990; Palla \& Stahler
1990, 1992).  In the present context, the masses of the forming stars
are sharply peaked about the mass scale $m_C \approx 2.3$.  For this
mass range, the most important source of luminosity arises from
infall, i.e., infalling material falls through the gravitational
potential well of the star and converts energy into photons.  The star
also generates internal luminosity which becomes important at
sufficiently high masses.  We can parameterize these two contributions
to find a luminosity versus mass relation of the form 
$${\widetilde L} = L_\ast / (1 L_\odot) = 70 \, \eta \, 
a_{35}^2 \, m \, + m^4 \, , \eqno(3.2)$$
where the first term arises from infall and the second term 
arises from internal luminosity (see Adams \& Fatuzzo 1996 for 
further discussion).  The efficiency parameter $\eta$ is the fraction
of the total available energy that is converted into photons. For
spherical infall, all of the material reaches the stellar surface and
$\eta \approx 1$.  For infall which includes rotation, some of the
energy is stored in the form of rotational and gravitational potential
energy in the circumstellar disk; we generally expect $\eta \sim
1/2$.  The sound speed $a$ determines the mass infall rate onto the
forming star/disk system according to ${\dot M} \sim a^3 / G$.  In
equation (3.2), we have written the sound speed in dimensionless form,
$a_{35} \equiv a/(0.35 \, {\rm km} \, \, {\rm s}^{-1})$.

We want to find a relationship between the distributions of the 
initial variables and the resulting distribution of stellar masses 
(the IMF).  For a given protostellar luminosity versus mass 
relationship, the semi-empirical mass formula 
can be written in the general form of a product of variables
$$M_\ast = \prod_{j=1}^n \alpha_j \, , \eqno(3.3)$$
where the $\alpha_j$ represent the variables which determine 
the masses of forming stars (the sound speed $a$, the rotation 
rate $\Omega$, etc., all taken to the appropriate powers). 
Each of these variables has a distribution $f_j (\alpha_j)$ 
with a mean value given by 
$$\ln \alphabar_j = \langle \ln\alpha_j \rangle =
\int_{-\infty}^\infty \ln\alpha_j \, f_j (\ln\alpha_j) \,
d\ln\alpha_j \, , \eqno(3.4)$$
and a corresponding variance given by  
$$\sigma_j^2 \, = \, \int_{-\infty}^\infty \, \xi_j^2
f_j (\xi_j) d\xi_j \,  . \eqno(3.5)$$

In the limit of a large number $n$ of variables, the composite 
distribution (i.e., the IMF) approaches a log-normal form. 
This behavior (see Adams \& Fatuzzo 1996; Zinnecker 1984)
is a direct consequence of the central limit theorem 
(e.g., Richtmyer 1978).  As a result, as long as a large number 
of physical variables are involved in the star formation process, 
the resulting IMF must be approximately described by a log-normal form. 
The departure of the IMF from a purely log-normal form depends on the 
shapes of the individual distributions $f_j$.  However, in the limit 
that the IMF can be described to leading order by a log-normal form, 
there are simple relationships between the distributions of the 
initial variables and the shape parameters $m_C$ and $\sigbar$ that 
determine the IMF. The mass scale $m_C$ is
determined by the mean values of the logarithms of the original
variables $\alpha_j$, i.e.,
$$m_C \equiv \prod_{j=1}^n \exp [ \langle \ln \alpha_j \rangle ]
\equiv \prod_{j=1}^n \alphabar_j \, , \eqno(3.6)$$
where we have defined $\alphabar_j$ =
$\exp [ \langle \ln \alpha_j \rangle ]$. 
The dimensionless shape parameter $\sigbar$ of the IMF determines 
the width of the stellar mass distribution and is given by the sum 
$$\sigbar^2 = \sum_{j=1}^n \sigma_j^2 \, . \eqno(3.7)$$

We can now use equations (3.6) and (3.7) to determine how 
the initial conditions for halo star formation must differ 
from star formation in present day molecular clouds. 
The results of the previous section show that the mass 
scale $m_C$ of the halo IMF must be larger than that of 
the present day IMF by a factor of $\sim10 - 20$.  This difference
implies that the mean values of the initial variables must be 
correspondingly larger so that the product (3.6) has the correct 
value. For example, we can consider the limit in which the effective 
sound speed is the most important physical variable. If we keep 
the mean values of all of the other variables the same as in the 
present day, then the mean sound speed must be in the range 
$\sim 0.70 - 0.90$ km/s ($\sim3$ times larger than present day) 
in order to obtain the mass scale $m_C = 2.3$.  For the case in 
which only thermal pressure contributes to the sound speed, this 
range of values corresponds to gas temperatures of $T_{\rm gas}$ 
$\sim 120  - 200$ K. Temperatures in this general range can be 
readily obtained in a zero metallicity environment through three body
cooling reactions (Palla, Stahler, \& Salpeter 1983; Lepp \& Shull
1983). Thus, the implied mass scale $m_C = 2.3$ is quite natural for
halo star formation.

The total dimensionless width of the IMF is constrained to be quite
small, with a maximum value $\sigbar \approx 0.44$.  This result
implies that the distributions of initial variables must be very
narrow.  In other words, the initial conditions for star formation in
the halo must be {\it very homogeneous}.  In order to quantify this
statement, we again consider the sound speed to be the most important
physical variable.  Suppose, for example, that the sound speed varies
by a factor $\cal F$.  For the relevant mass range, the SEMF implies
that $M_\ast \sim a^3$ (where we have combined equations [3.1] and
[3.2]).  Thus, the contribution of the variation in the sound speed to
the total variance is given by $\sigma_a^2 = (3 \ln {\cal F})^2$.  If
the variance in the sound speed accounts for the entire variance in
the mass distribution (i.e., we assume that $\sigma_a^2 = \sigbar^2$),
then we can solve for the factor $\cal F$ = 1.16.  In other words, the
effective sound speed can only vary by 16\% in the primordial fluid
that produced this generation of stars. This result, in turn,
corresponds to an allowed temperature variation of only 32\%. If we
allow the other parameters in the SEMF to vary as well, then the sound
speed and temperature are constrained to vary by even less than these
amounts.

Before leaving this section, we briefly consider the idea that 
stellar masses are determined by the Jeans mass 
$$M_J \equiv {4 \pi \over 3} \Bigl( {\pi a^2 \over G} \Bigr)^{3/2} 
\rho^{-1/2} \, . \eqno(3.8)$$ 
For the sound speed $a = 0.90$ km/s derived above (this value is
also consistent with primordial cooling calculations), the Jeans mass
is given by $M_J = 3.5 \times 10^5 M_\odot \, n^{-1/2}$, where $n$ is
the number density of the gas.  Thus, to obtain the characteristic mass 
scale $m_C = 2.3$ required for the IMF, the number density must be
extremely large: $n = 2 \times 10^{10}$ cm$^{-3}$.  Since this value
is many orders of magnitude larger than any expected density at the
epoch of star formation in the galactic halo, the idea that the Jeans
mass determines the mass scale of forming stars is essentially ruled
out.
 
\bigskip 
\centerline{\bf 4. THE WHITE DWARF LUMINOSITY FUNCTION} 
\medskip 

In this section, we consider the ramifications of the posited halo
white dwarfs on the observed luminosity function.  In particular, we
show that in order for the MACHO Collaboration's lensing result to be
consistent with the present-day white dwarf luminosity function, the
age of the majority of the halo dwarfs must exceed $\sim$16 Gyr.

The nature of the local white dwarf luminosity function has been
the focus of considerable prior effort. Schmidt (1959) was the first
to point out that the star formation history of the galactic disk is
written into the current observed population of cooling white
dwarfs. Specifically, he noted that there should be no white dwarfs
whose cooling times exceed the disk age, thus predicting a drastic
falloff in number density at a specific luminosity.  These ideas were
quantified and extended by D'Antona and Mazzitelli (1978).

Liebert et al. (1979) demonstrated the existence of an abrupt falloff
in the observed number of white dwarf stars below a luminosity
$\log_{10} (L/L_\odot) \approx$ --4.5 (see also Liebert 1980). 
In a pioneering effort, Winget et al. (1987) computed theoretical
luminosity functions from the results of white dwarf cooling theory,
and compared these functions with the observational data; they obtained
a disk age of 9.3 $\pm$ 2.0 Gyr. More detailed studies of the white
dwarf luminosity function confirmed this estimate of the disk age and
showed that the result is remarkably robust when confronted with 
variations in the necessary input parameters (e.g., Iben \& Laughlin 
1989; Yuan 1989; Noh \& Scalo 1990; and Wood 1992). 

Several physical relations are needed to construct a model luminosity
function for a population of cooling white dwarfs. The cooling curves 
themselves are of primary importance. Here, we ignore the complications 
brought on by chemical composition, and we assume that the cooling time 
is a function of the white dwarf mass $m_{WD}$ and luminosity $\ell$:  
$$t_{cool}=t_{c}(\ell,m_{WD}). \eqno(4.1)$$
Further, we assume a relation between the mass $m_{WD}$ of the white 
dwarf and the mass $m$ of its main-sequence progenitor:
$$m_{WD} = m_{WD} (m) \, . \eqno(4.2)$$
We also require a relation for the nuclear burning lifetime of 
the progenitor as a function of mass, 
$$t_{MS} = t_{\rm evol} (m), \eqno(4.3)$$
as well as the IMF $dN/dm$.  If we assume that the main-sequence 
precursors to the current lensing population all formed at a single 
time $\tau_H$, each luminosity interval $d\ell$ centered around a 
particular luminosity $\ell$ is populated by white dwarfs of mass 
$m_{WD}$ which satisfy the relation: 
$$t_{\rm evol} (m) + t_{c}(\ell,m_{WD}) = \tau_H \, . \eqno(4.4)$$

Differentiating equations (4.2) and (4.4), we can write down the 
form of the luminosity function due to the present day population 
of halo white dwarfs: 
$${dn\over{dM_{\rm bol}}}={\ln 10 \over{2.5}}
{ \phi\, (dN/dm) \, \, t_{c}\, \, (\partial \log_{10} t_{c}
/\partial \log_{10} \ell)_{m_{WD}} \over 
{(dt_{\rm evol}/dm) + (t_{c}/m_{WD}) 
(\partial \log_{10} t_{c} / \partial \log_{10} m_{WD})_{\ell}
(d m_{WD} / dm)}} \, , \eqno(4.5)$$
where $\phi$ is the normalization factor required to produce the
requisite number density of white dwarfs in the solar neighborhood
implied by the MACHO result. In keeping with the usual
convention, we have written the luminosity function in terms of 
the differential magnitude $dM_{\rm bol}$ rather than the 
differential luminosity $d\ell$.  Further details regarding
the derivation of equation (4.5) are given in Iben \& Laughlin (1989).

If we assume that the stars that gave rise to the white dwarfs formed
over a period of time (rather than in a single burst), then the star
formation rate $\phi(t)$ must be incorporated into a continuous
formulation of the luminosity function (see Noh \& Scalo (1990) 
for a rigorous derivation): 
$${dn \over dM_{\rm bol}} = {\ln 10 \over {2.5}} \ell 
\int_{M_{1}}^{M_{2}} \phi \bigl[ t(m,\ell) \bigr] \, 
{dN \over{dm}} \, \Bigl( {\partial t_{c} \over 
{\partial \ell}} \Bigr)_{m_{WD}} dm \, . \eqno(4.6)$$

As mentioned above, the previous efforts aimed at synthesizing white
dwarf luminosity functions have all shown that the end results are
relatively insensitive to most variations in the input physics. Our
approach is therefore to define a standard model, explore its
consequences, and then briefly discuss the secondary effects brought
about by changes in the input parameters.

In light of the MACHO result, the total luminosity function of white
dwarfs in the solar neighborhood should incorporate members from both
the disk and the halo populations. Our synthesis program accounts for
the disk population through numerical integration of equation (4.6),
and for the halo population through evaluation of equation (4.5).  
For the disk contribution to the luminosity function, we adopt a 
standard model which is rather similar to the one introduced by 
Wood (1992).  In particular, we assume the following: 

\item{[1]} The age of the disk $\tau_{\rm disk}$ = 9.0 Gyr.

\item{[2]} A constant star formation rate $\phi(t)$ = {\sl constant}
= 5.0 $\times 10^{-13}$ pc$^{-3}$ yr$^{-1}$. This value represents the 
star formation rate required to produce the current observed white dwarf
density of 3.0 $\times 10^{-3}$ pc$^{-3}$ for luminosities in the range 
$\log_{10} (L/L_{\odot}) >$ $-4.5$.

\item{[3]} A Salpeter IMF for the {\it disk} population: 
$dN/dm \propto m^{-2.35}$ (Salpeter 1955). 

\item{[4]} White dwarf progenitor lifetimes are given by the formula, 
$$\log_{10} t_{\rm evol} = 9.921 - 3.6648 (\log_{10} m) + 
1.9697(\log_{10} m)^{2} - 0.9369(\log_{10} m)^{3} \, . \eqno (4.7)$$
This result is taken from Iben \& Laughlin (1989) who
obtained the relation by extracting main sequence lifetimes from the 
stellar evolution calculations of a number of different authors.

\item{[5]} The standard relationship between progenitor mass and 
white dwarf mass from Wood (1992).  This relation is given in 
equation (2.10). 

The most important element in a model luminosity function is the set
of mass-dependent white dwarf cooling curves. Unfortunately, however,
the theory of white dwarf cooling is both complicated and uncertain.  
In our calculations, we have used the cooling sequences of Winget et al. 
(1987), which span masses of $m$ = 0.4, 0.6, 0.8, and 1.0 down to
luminosities of $\log_{10}$ $(L/L_{\odot})$ = --5.0.  These curves are
based on white dwarf models composed of pure carbon.  The interiors of
real white dwarfs are believed to be an admixture of carbon and
oxygen, and the envelopes are either helium, or, in the case of DA
dwarfs, helium and hydrogen. However, as stressed by Winget {et al.}
(1987), the adjustments to the cooling times produced by the
hydrogen-helium envelopes and the oxygen admixtures in the cores tend
to cancel one another out. Hence the pure carbon models should produce
a reasonable approximation to the actual cooling curves.

Because the halo population has had a long time to cool off, 
the cooling curves (Winget {et al.} 1987) must be extrapolated to 
luminosities below $\log_{10} (L/L_{\odot}) = -5.0$.  As discussed 
by Wood (1992), white dwarfs dimmer than $\log_{10} (L/L_{\odot})  
\approx -5.0$ are almost entirely crystalized. 
As a result, they are in the {\it Debye} regime and hence 
suffer from very low heat capacities, and will cool to invisibility 
within a finite time. Therefore, naive linear extrapolations 
of the cooling curves below $\log_{10} (L/L_{\odot}) = -5.0$ 
tend to severely overestimate cooling times, which in turn adversely 
affect the number density estimates at the faint end of the luminosity 
function. To more accurately account for the effects of the Debye 
regime, we extend each cooling curve down to the value  
$\log_{10} (L/L_{\odot}) = -6.25$ using a prescription very 
similar to the one advocated by Wood (1992):
$$t-t_{0} = A \Bigl[ 1 - \bigl( {L \over L_{0} } 
\bigr) \Bigr] \, . \eqno(4.8)$$
In equation (4.8), the reference values $t_{0}$ and $L_{0}$ are 
the age and luminosity of the $\log_{10} (L/L_{\odot}) = -5.0$ model
of each mass sequence. The constant $A$ is determined using the last
two tabulated points in each of the cooling curves (from Winget et
al. 1987).

Derivatives of the cooling curves at arbitrary values $(m_{WD},\ell)$ 
with respect to both mass and luminosity are required to evaluate 
equations (4.5) and (4.6).  Using centered differencing, we compute 
the derivatives at each of the tabulated points. Values for $t_{c}$, 
$\partial \log_{10} t_{c}/\partial \log_{10} \ell$, and 
$\partial \log_{10} t_{c}/ \partial \log_{10} m_{WD}$ 
are then obtained by using bicubic splines.  Interpolations are 
performed in the logarithm of all variables.

After being extended and interpolated, the Winget et al. (1987)
cooling curves can be used to model the luminosity functions of both
the halo and the disk.  To build halo white dwarf luminosity
functions, we again use the white dwarf progenitor lifetime given by
equation (4.7) and the conversion relation of equation (2.10). The
normalization factor in equation (4.5) is set at $\phi = 4.43 \times
10^{-3}$.  This value corresponds to the total number of halo white
dwarfs per pc$^{3}$ at the solar circle, if one assumes an average
white dwarf mass of $m_{WD}$ = 0.63, a total halo density given by
equation [5.4] below, and a total white dwarf contribution to the halo
mass of 25\%.  Coincidentally, this value for $\phi$ yields a halo white
dwarf number density which is slightly greater than the number density 
of {\it known} white dwarfs in the solar neighborhood (nearly all of 
which are certainly members of the warmer, brighter disk population).
As our standard case, we adopt the log-normal IMF determined in \S 2,
with values $m_C$ = 2.3 and $\sigbar = 0.44$. 

Using this aggregate of input physics, we compute the composite
disk+halo luminosity functions.  The result for our standard IMF is
displayed in Figure 4a. For comparison purposes, we have plotted the
observed local white dwarf luminosity function (from Liebert, Dahn, \& 
Monet 1988).  The three low luminosity points are represented by error
boxes (following Wood 1992).  In this representation, theoretical
models which pass through all three boxes are considered to be
consistent with the observed data.

The luminosity function for the disk population alone is indicated 
by the dotted line. The fact that this dotted line can only be 
distinguished from the composite (disk+halo) luminosity functions at
low luminosities indicates that the disk population is entirely
responsible for producing existing white dwarfs brighter than
$\log_{10} (L/L_{\odot}) \approx -4.0$.  The overall quality of our 9
Gyr disk luminosity function fit to the data is quite good. This
finding is in concordance with the previous studies of the disk white
dwarf population.  As a side note, the marginal excess of stars
observed at $\log_{10} (L/L_{\odot}) \approx -2.0$, has been attributed
to a burst of star formation activity (Noh \& Scalo 1990) which is
believed to have occurred $\sim$6 Gyr ago. 

Our assumption of an IMF with the shape parameters $m_C$ = 2.3 and
$\sigbar = 0.44$, combined with the unchallenged existence of an
observed decline in the white dwarf luminosity function, essentially
rules out halo populations which are less than $\sim$16 Gyr old.
Although such an age is not currently attractive from a cosmological
standpoint, it stands in good agreement with the ages of the oldest
globular clusters, which are thought to lie in the range 15--18 Gyr
(e.g., Vandenberg 1983).  Furthermore, the dating method provided 
by white dwarf cooling is only obliquely dependent on the stellar 
evolution calculations which determine the globular cluster
ages. Variations in the main sequence progenitor lifetime relation
have only a moderate effect on the age determination of the halo 
dwarf population. 

Although the MACHO result suggests that roughly half of the halo mass
resides in white dwarfs, we have argued that this fraction is perhaps
more likely to be on the order of 25\%, as required by mass
limitations on a single generation of progenitor stars.  Only a
sustained epoch of halo star formation involving several generations
can account for white dwarfs composing 50\% of the halo. Nevertheless,
a white dwarf halo population which has twice our assumed number
density, and a time-of-formation spread of several billion years would
not change the mean age estimate significantly.  In order to produce
the observed falloff in the luminosity function, the majority of the
white dwarfs still must be older than $\sim$16 Gyr.

The bounds of the lowest luminosity error box are primarily determined
by two very dim white dwarfs.  A measurement of the temperature
$T_{\rm eff}$ for one of these objects (LP 701-29) indicates that this
star has a radius $R \approx 0.01 R_{\odot}$ and a mass $m \approx$
0.6 (Kapranidis \& Liebert 1986).  These values are slightly odd if
the object is a member of the 9 Gyr disk population. For our choice of
input physics, the {\it least} massive disk star whose white dwarf has
had time to cool to the LP 701-29 luminosity of $\log_{10}
(L/L_{\odot}) \approx -4.5$ has a mass of $m$ = 4.0, and a remnant
white dwarf mass of $m_{WD}$ = 0.72.  Winget et al. (1987) quote a
dominant white dwarf mass of $m_{WD}$ = 0.80 at the shortfall.  On
the other hand, if LP 701-29 belongs to a 16 Gyr old halo population,
then its progenitor, whose dwarf now lies at the extreme bright end of
the halo distribution, would have had a mass of $m$ = 1.04, 
indicating a white dwarf mass of $m_{WD}$ = 0.55. 

Figures 4b and 4c chart the sequences of composite white dwarf
luminosity functions which result from allowed variations in the 
shape parameters of the IMF. In Figure 4b, the shape parameters 
are $m_C$ = 2.5 and $\sigbar = 0.30$, whereas in Figure 4c, the shape 
parameters are $m_C$ = 3.0 and $\sigbar = 0.20$. Both of these alternate
distributions have narrower peaks around higher masses, and hence
produce considerably fewer stars of solar mass. Hence, their 
preponderance of high mass stars makes these IMFs less desirable 
in light of the already serious white dwarf efficiency problem 
outlined in \S 2. Nevertheless, even for these distributions, 
it is clear that the halo population must be
considerably older than the disk in order to account for the observed
paucity of white dwarfs at luminosities near $\log_{10} (L/L_{\odot})
\approx -4.5$.  In each diagram, the heavy solid line corresponds to
the luminosity function that most reasonably fits the observed data.

\newpage 
\centerline{\bf 5. THE BACKGROUND RADIATION FIELD} 
\medskip 

In this section, we calculate the radiation signature of a
galactic halo composed of white dwarfs (we follow the general 
formulation of Adams \& Walker 1990). We first assume that at 
any given spatial point in the halo the mass distribution and
properties of the white dwarf population are the same.
The differential flux density $d F_\nu$ at frequency $\nu$
received at the earth from a particular point in the galactic
halo in a particular direction is given by
$$d F_\nu = \gnu \Bigl( {\Omega_T \over 4 \pi} \Bigr)
\, \rho  \, ds , \eqno(5.1)$$
where $\rho$ is the mass density of the halo,
$\Omega_T$ is the angular size of the observing beam, and
$ds$ is the line element along the given line of sight.
The specific luminosity $\gnu$ is defined such that
$\gnu/4 \pi r^2$ is the flux density emitted at frequency $\nu$
per unit mass.  The total observed flux density $F_\nu$ is obtained
by integrating equation (5.1) along the line of sight,
$$F_\nu = \gnu \Bigl( {\Omega_T \over 4 \pi} \Bigr)
\int \rho [r(s)] \, ds \equiv
\gnu \Bigl( {\Omega_T \over 4 \pi} \Bigr)
\rho_0 R \, \, {\cal I}(b, \ell) ,  \eqno(5.2)$$
where $R$ is the distance from the sun to the galactic center,
$\rho_0$ is a fiducial value of the density of the galactic halo
(see eq. [5.4]), and where ${\cal I}(b, \ell)$ is a dimensionless
integral (see eq. [5.5]) which depends on the viewing angle
(given in galactic coordinates).  Notice that we implicitly assume
that the beam is sufficiently small to consider only a single line 
of sight in the integral (rather than integrating over the beam).  
The determination of the flux signature divides cleanly into three 
separate components (Adams \& Walker 1990): the radiative component 
$\gnu$ which depends only on the properties of the white dwarf
population, the line of sight integral which depends on the density
distribution of the galactic halo, and the solid angle $\Omega_T$
which depends on telescope properties.  In the following discussion,
we specify the properties of the white dwarf population and the
density distribution of the galactic halo and then calculate the
radiative flux.

The specific luminosity $\gnu$ can be written
$$\gnu = {
\int dm_{WD} (dN/dm_{WD}) \, 4 \pi R_\ast^2 (m_{WD}) \,
\pi B_\nu [T_\ast (m_{WD})]
\over \int \, m_{WD} \, dm_{WD} (dN/dm_{WD}) } \, , 
\eqno(5.3{\rm a})$$
where the $dN/dm_{WD}$ is the distribution of masses of the 
white dwarfs which make up the halo.  In the limit that the 
mass distribution is a delta function, the specific luminosity 
reduces to the simple form 
$$\gnu = 4 \pi^2 R_\ast^2 B_\nu[T_\ast]/m_{WD} 
\, . \eqno(5.3{\rm b})$$  
The white dwarf radius $R_\ast(m_{WD})$ depends only on the mass 
$m_{WD}$ and can be obtained from standard white dwarf models 
(e.g., Shapiro \& Teukolsky 1983).  For a given mass and age, the 
luminosity of a given white dwarf (and hence its stellar temperature 
$T_\ast$) can be determined from the considerations 
outlined in the previous section. 

Next, we specify the mass density $\rho$ of the galactic halo.  
We use a simple halo model with a spherically 
symmetric density distribution of the form 
$$\rho (r) = \rho_0 \, 
{ {\varpi^2} \over {\varpi^2 + r^2}} , \eqno(5.4)$$
where $r$ is the radial distance from the galactic center,
$\rho_0$ $\approx$ 1.3 $\times$ $10^{-23}$ g cm$^{-3}$, and 
$\varpi$ $\approx$ 2 kpc (Binney \& Tremaine 1988; 
Bahcall \& Soneira 1980). In order to determine the halo emission, 
we must evaluate the non-dimensional integral
$${\cal I}(b, \ell) = {1 \over \rho_0 R} \, \int_0^\infty ds \,
\rho[r(s)] \, \, = \int_0^\infty {ds \over R} \, \,
{\varpi^2 \over \varpi^2 + r^2 } , \eqno(5.5)$$
where $R$ = 8.5 kpc is the distance to the galactic center.
In order to specify the direction angles,
we use the galactic coordinates $(b, \ell)$
which correspond to a spherical coordinate system centered
on the sun.  The $z$-direction coincides with the direction
of the North galactic pole, and the $x$-direction (the
zero of the galactic longitude $\ell$) coincides with the
direction of the galactic center. Notice that the galactic
latitude $b$ is measured from the galactic equator (the
plane of the disk) rather than from the galactic pole.
If we define the quantities
$$\alpha \equiv \varpi / R \qquad {\rm and} \qquad
\mu \equiv \cos b \, \cos\ell , $$
then the integral $\cal I$ can be evaluated to obtain the form 
$${\cal I}(b, \ell) =
{\alpha^2 \over [1 + \alpha^2 - \mu^2]^{1/2} }
\Biggl\{ {\pi \over 2} +
\tan^{-1} {\mu \over [1 + \alpha^2 - \mu^2]^{1/2} }
\Biggr\} . \eqno(5.6)$$
Notice that the radiation field emitted by 
galactic halos exhibits a well defined angular dependence. 
This modulation of the radiative signature can help distinguish 
observations of the halo from other possible sources of radiation 
(see also Gunn et al. 1978; Kephart \& Weiler 1987;
Adams \& Walker 1990).

Using the above formulation, the IMF shape parameters given 
by equation (2.9), and the white dwarf mass versus progenitor 
mass relationship (2.10), we can determine the radiation field 
produced by a halo filled with white dwarfs.  The result is 
shown in Figure 5 for assumed halo ages of $\tau_H$ = 10, 12, 
14, and 16 Gyr. We have normalized the curves such that {\it all} 
of the halo mass is in the form of white dwarfs.  These curves 
should thus be scaled by the assumed mass fraction of white dwarfs 
in the halo.  However, even for the brightest possible white 
dwarf halo (age of 10 Gyr and mass fraction of unity), the 
background radiation field (with brightness $I_\nu \sim 100$ Jy/ster) 
is safely below current observational limits.  

It is possible that future satellite missions (e.g., the SIRTF project
currently being developed by NASA) can achieve sufficient sensitivity
to either detect or rule out the brightest of these white dwarf halos.
The SIRTF satellite is expected to have a sensitivity of $\sim 2000$
Jy/ster for each resolution element in its array camera (this value
corresponds to a 3 $\sigma$ detection with an integration time of one
hour and a wavelength in the range 2--6 $\mu$m).  Thus, measuring even
the brightest possible white dwarf halo ($I_\nu \sim 100$ Jy/ster)
would require very long time integrations and co-adding of resolution
elements.

\newpage 
\centerline{\bf 6. SUMMARY AND DISCUSSION} 
\medskip 

In this paper, we have discussed the implications of a galactic 
halo which contains a substantial mass fraction of white dwarfs. 
Our specific results can be summarized as follows: 
 
\item{\bf [1]} We have shown that the IMF of the initial stellar 
population is very highly constrained.  As a result, the vast 
majority of stars from this population must lie in the mass 
range $1 < m < 8$ required for white dwarf production.  
This IMF is thus very different from the present day IMF (see 
Figures 1 and 2).  The IMF shape parameters that saturate the 
constraints correspond to a mass scale $m_C \approx 2.3$
and a dimensionless width $\sigbar \approx 0.44$.

\item{\bf [2]} White dwarfs cannot make up the entire mass of the 
galactic halo if only a single stellar generation produced the 
compact objects.  The white dwarf efficiency factor ${\cal E}_{WD}$ 
$\approx$ 0.24 for the IMF described in item [1].  The maximum 
possible white dwarf efficiency factor is ${\cal E}_{WD} \sim$ 0.5, 
although it is extremely unlikely that this limit can be realized 
in practice.  In addition, the large amount of gas left over from 
the process poses severe problems for a large mass fraction of 
white dwarfs in the galactic halo. 

\item{\bf [3]} We have used current theoretical IMF developments in
conjunction with the above limits on the IMF to constrain the initial
conditions for star formation at the epoch of halo star formation.
The initial conditions must be much more homogeneous that those which
lead to the present day IMF.  In addition, the mean values of the
parameters which determine stellar masses must be skewed toward higher
values.  For example, the total effective sound speed (which
represents the most important physical variable in the problem) must
have a mean value of $a \approx$ 0.90 km/s, about a factor of three
larger than the value implied by the present day IMF. This sound speed
corresponds to an effective temperature of $T_{\rm gas}$ = 200 K.

\item{\bf [4]} The white dwarf population in the galactic halo 
dominates the total white dwarf luminosity function for low 
luminosities $\log_{10} (L/L_\odot) < -4.5$. The shape and 
amplitude of the luminosity function is a rather sensitive 
function of the age of the halo population.  As a result, 
existing limits on the white dwarf luminosity function imply 
that the age of the halo population must be larger than 
$\sim 16$ Gyr.  Furthermore, if white dwarfs make up a 
substantial fraction of the halo mass, then the white dwarf 
luminosity function must turn up at luminosities just fainter 
than the current limits (see Figure 4). 

\item{\bf [5]} We have determined the radiative signature of a 
galactic halo with a substantial mass fraction of white dwarfs 
(see Figure 5). The radiation field from such a halo is very faint
and hence below current observational limits, but is almost 
detectable with future satellite missions. 

The net results of microlensing experiments indicate that a
substantial fraction of the halo mass resides in some type of low mass
stellar objects.  The only two reasonable candidates for these objects
are either brown dwarfs or white dwarfs.  The brown dwarf hypothesis
implies a halo IMF which is sharply peaked about a very low mass
scale, $m_C < 0.05$.  The white dwarf hypothesis implies a halo IMF
which is sharply peaked about a larger mass scale $m_C = 2 - 3$.  
For comparison, the present day IMF has a mass scale $m_C = 0.1 - 0.2$ 
intermediate between these values and a much wider distribution (i.e., 
a much larger value of $\sigbar$).  We argue that current theoretical
work on star formation and the IMF strongly favors the white dwarf
scenario (over the brown dwarf scenario).  If stars play a role in
determining their own masses through the action of stellar winds, then
larger sound speeds (larger temperatures) lead to the production of
higher mass stars.  Since the sound speed (temperature) is expected to
be higher during the epoch of halo star formation, the white dwarf
scenario is more natural from the point of view of star formation
theory.  Furthermore, a halo with a considerable fraction of its mass
in white dwarfs appears to require that the universe is older than at
least 15 Gyr and provides a specific testable prediction: there should
be a dramatic upturn in the white dwarf luminosity function at
luminosities only slightly dimmer than those at which white dwarfs are
currently being detected.

The results of this paper indicate that the most problematic issue for
white dwarf halos is the large amount of gas left over from the
production process.  In other words, as shown in \S 2, the white dwarf
efficiency is quite low.  This leftover gas must either be locked up
in other forms of baryonic dark matter or stripped out of the galaxy
and into the intergalactic medium.  If the observational result
indicating a large mass fraction of white dwarfs is confirmed, then
future theoretical work should concentrate on this important issue.

\vskip0.5truein
\centerline{Acknowledgements}

We would like to thank Carl Akerlof, Gus Evrard, and Joe Mohr 
for stimulating astrophysical discussions. 
This work was supported by an NSF Young Investigator
Award, NASA Grant No. NAG 5-2869, and by funds from the Physics
Department at the University of Michigan.

\newpage 
\centerline{\bf REFERENCES} 
\medskip 

\par\pp
Abramowitz, M., \& Stegun, I. A. 1970, Handbook of Mathematical 
Functions (New York: Dover) 

\par\pp 
Adams, F. C. 1990, {\sl ApJ}, {\bf 363}, 578

\par\pp
Adams, F. C., \& Fatuzzo, M. 1996, {\sl ApJ}, in press 

\par\pp
Adams, F. C., \& Shu, F. H. 1986, {\sl ApJ}, {\bf 308}, 836

\par\pp
Adams, F. C., \& Walker, T. P. 1990, {\sl ApJ}, {\bf 359}, 57 

\par\pp
Alcock, C. et al. 1993, {\sl Nature}, {\bf 365}, 621

\par\pp
Aubourg, E. et al. 1993, {\sl Nature}, {\bf 365}, 623

\par\pp
Bahcall, J. N., \& Soneira, R. M. 1980, {\sl Ap. J. Suppl.},
{\bf 44}, 73

\par\pp
Bahcall, J. N., Flynn, C., Gould, A., Kirhakos, S. 1994,
{\sl ApJ}, {\bf 435}, L51 

\par\pp
Bennet, D. et al. 1996, {\sl Bull. AAS}, {\bf 28}, 47.07 

\par\pp
Binney, J., \& Tremaine, S. 1987, {Galactic Dynamics} 
(Princeton: Princeton University Press) 

\par\pp
Charlot, S., \& Silk, J. 1995, {\sl ApJ}, {\bf 445}, 124  

\par\pp
D'Antona, F., \& Mazzitelli, I. 1978, {\sl A \& A}, {\bf 66}, 453

\par\pp
Elmegreen, B. G., \& Mathieu, R. D. 1983, {\sl MNRAS}, {\bf 203}, 305

\par\pp
Graff, D. S., \& Freese, K. 1996, {\sl ApJ}, {\bf 456}, L49 

\par\pp
Gunn, J. E., Lee, B. W., Lerche, I., Schramm, D. N., \& Steigman, G.
1978, {\sl ApJ}, {\bf 223}, 1015

\par\pp
Hegyi, D. J., \& Olive, K. A. 1983, {\sl Phys. Lett.}, {\bf 126B}, 28 

\par\pp
Hegyi, D. J., \& Olive, K. A. 1986, {\sl ApJ}, {\bf 303}, 56 

\par\pp
Hegyi, D. J., \& Olive, K. A. 1989, {\sl ApJ}, {\bf 346}, 648 

\par\pp
Iben, I., \& Laughlin, G. 1989, {\sl ApJ}, {\bf 341}, 312

\par\pp
Kapranidis, S., \& Liebert, J. 1986, {\sl ApJ}, {\bf 305}, 863 

\par\pp
Kephart, T. W., \& Weiler, T. J. 1987,
{\sl Phys. Rev. Lett.}, {\bf 58}, 171

\par\pp
Larson, R. B. 1973, {\sl MNRAS}, {\bf 161}, 133

\par\pp 
Lepp, S., \& Shull, J. M. 1983, {\sl ApJ}, {\bf 270}, 573 

\par\pp 
Liebert, J. 1980, in IAU Colloquium 53, {White Dwarfs and Variable 
Degenerate Stars}, ed. H. M. Van Horn \& V. Weidemann (Rochester:
University of Rochester), p. 146

\par\pp 
Liebert, J., Dahn, C. C., Gresham, M., \& Strittmatter, P.A. 
1979, {\sl ApJ}, {\bf 233}, 226

\par\pp
Liebert, J., Dahn, C. C., \& Monet, D. G. 1988, {\sl ApJ}, {\bf 332}, 891

\par\pp
MACHO Collaboration 1996, New York Times, 17 January 

\par\pp
Miller, G. E., \& Scalo, J. M. 1979, {\sl ApJ Suppl.}, {\bf 41}, 513

\par\pp
Noh, H.-R., \& Scalo, J. 1990, {\sl ApJ}, {\bf 352}, 605

\par\pp
Palla, F., Salpeter, E. E., \& Stahler, S. W. 1983, {\sl ApJ}, 
{\bf 271}, 632 

\par\pp
Palla, F., \& Stahler, S. W. 1990, {\sl ApJ}, {\bf 360}, L47

\par\pp
Palla, F., \& Stahler, S. W. 1992, {\sl ApJ}, {\bf 392}, 667 

\par\pp
Richtmyer, R. D. 1978, {Principles of Advanced Mathematical Physics}
(New York: Springer-Verlag)

\par\pp
Ryu, D., Olive, K. A., \& Silk, J. 1990, {\sl ApJ}, {\bf 353}, 81 

\par\pp
Salpeter, E. E. 1955, {\sl ApJ}, {\bf 121}, 161 

\par\pp
Scalo, J. M. 1986, {\sl Fund. Cos. Phys.}, {\bf 11}, 1

\par\pp
Shapiro, S. L., \& Teukolsky, S. A. 1983, {Black Holes, 
White Dwarfs, and Neutron Stars: The Physics of Compact Objects}, 
(New York: Wiley) 

\par\pp 
Schmidt, M. 1959, {\sl ApJ}, {\bf 129}, 243

\par\pp
Shu, F. H. 1977, {\sl ApJ}, {\bf 214}, 488 

\par\pp
Shu, F. H., Adams, F. C., \& Lizano, S. 1987,
{\sl A R A \& A}, {\bf 25}, 23 

\par\pp
Shu, F. H., Lizano, S., \& Adams, F. C. 1987, in Star Forming
Regions, IAU Symp. No. 115, ed. M. Peimbert \& J. Jugaku
(Dordrecht: Reidel), p. 417 

\par\pp
Stahler, S. W., Shu, F. H., \& Taam, R. E. 1980, {\sl ApJ},
{\bf 241}, 63

\par\pp
Vandenberg, D. A. 1983, {\sl ApJ Suppl.}, {\bf 51}, 29

\par\pp 
Winget, D.E., Hansen, C.J., Liebert, J., Van Horn, H. M., 
Fontaine, G., Nather, R. E., Kepler, S. O., \& Lamb, D. Q. 1987, 
{\sl ApJ}, {\bf 315}, L77 

\par\pp
Wood, M. A. 1992, {\sl ApJ}, {\bf 386}, 529

\par\pp
Yuan, J. W. 1989, {\sl A \& A}, {\bf 224}, 108

\par\pp
Zinnecker, H. 1984, {\sl MNRAS}, {\bf 210}, 43 

\par\pp
Zinnecker, H. 1985, in Birth and Infancy of Stars, eds.
R. Lucas, A. Omont, \& R. Stora (Amsterdam: North Holland), p. 473

\newpage 
\centerline{\bf FIGURE CAPTIONS} 
\medskip 

\medskip 
Figure 1. The derived IMF at the epoch of star formation in the halo
(shown as the solid curve).  This IMF saturates the constraints of 
\S 2 and has shape parameters $m_C$ = 2.3 and $\sigbar$ = 0.44.  
A fit to the present day IMF (from Miller \& Scalo 1979) is
also shown for comparison (dashed curve).

\medskip 
Figure 2. Allowed values for the shape parameters in the halo IMF.  
The allowed region of parameter space is shown as the hatched portion 
in the lower $m_C$-$\sigbar$ plane. The square symbol in the upper 
left part of the diagram shows the location corresponding 
to the present day IMF. 

\medskip 
Figure 3. Mass distribution of white dwarfs in the halo derived from the 
halo star IMF (solid curve).  The mass distribution of white dwarfs in 
the disk (derived from the present day IMF) is also shown for comparison 
(dashed curve). 

\medskip 
Figure 4. Luminosity function for white dwarfs, including both the 
disk population (with age 9 Gyr) and the halo population (with ages
varying in the range 10 -- 20 Gyr). 
(a) White dwarf halo population arising from an IMF with our standard 
values of the shape parameters $m_C$ = 2.3 and $\sigbar$ = 0.44. 
(b) White dwarf halo population arising from an IMF with 
shape parameters $m_C$ = 2.5 and $\sigbar$ = 0.30. 
(c) White dwarf halo population arising from an IMF with 
shape parameters $m_C$ = 3.0 and $\sigbar$ = 0.20. 

\medskip 
Figure 5. Radiative signature of a galactic halo composed 
of white dwarfs.  Curves show the background infrared radiation 
field for a halo composed of white dwarfs with (total) ages 
$\tau_H$ = 10, 12, 14, and 16 Gyr (from top to bottom). 
All curves use a log-normal IMF (with shape parameters 
$m_C$ = 2.3 and $\sigbar$ = 0.44) and the white dwarf/progenitor 
mass relationship of equation (2.10).

\bye